\documentclass[a4paper,11pt]{article}
\usepackage{pos}
\usepackage{listings}

\lstdefinestyle{mystyle}{
	backgroundcolor=\color{darkgray}, 
	basicstyle=\color{white}\small, 
	stringstyle=\rmfamily\itshape,
	language=Python,
	commentstyle=\color{yellow},
}

\title{Lyncs-API: a Python API for Lattice QCD applications}
\ShortTitle{Lyncs-API}

\author*[a]{Simone Bacchio}
\author[a]{Jacob Finkenrath}
\author[a]{Christodoulos Stylianou}

\affiliation[a]{Computation-based Science and Technology Research Center, The Cyprus Institute, Cyprus}

\emailAdd{s.bacchio@cyi.ac.cy}

\abstract{We present Lyncs-API, a Python API for Lattice QCD applications currently under development. Lyncs aims to bring several widely used libraries for Lattice QCD under a common framework. Lyncs flexibly links to libraries for CPUs and GPUs in a way that can accommodate additional computing architectures as these arise, achieving performance-portability for the calculations while maintaining the same high-level workflow. Lyncs distributes calculations using \textit{Dask} and \textit{mpi4py}, with bindings to the libraries automatically generated by \textit{cppyy}. While Lyncs is designed to allow linking to multiple libraries, we focus on a set of targeted packages that include \textit{DDalphaAMG}, \textit{tmLQCD}, \textit{QUDA} and \textit{c-lime}. More libraries will be added in the future. We also develop generic-purpose tools for facilitating the usage of Python in Lattice QCD and HPC in general. The project is open-source, community-oriented and available on Github.
}

\FullConference{%
 The 38th International Symposium on Lattice Field Theory, LATTICE2021
  26th-30th July, 2021
  Zoom/Gather@Massachusetts Institute of Technology
}

\newcommand{\lyncs}[1]{\href{https://github.com/Lyncs-API/lyncs.#1}{\texttt{lyncs\_#1}}}
\newcommand{\Lyncs}{\href{https://github.com/Lyncs-API/lyncs}{\texttt{lyncs}} }
\newcounter{packages}
\newcommand{\package}[1]{
	\href{https://pypi.org/project/#1}{\texttt{#1}}\refstepcounter{packages}\label{package:#1}
}
\newcommand{\pkg}[1]{
	\hyperref[package:#1]{\texttt{#1}}
}
\newcommand{\fhref}[2]{\href{#1}{#2}\footnote{\url{#1}}}

\begin{document}
\tableofcontents
\maketitle

\section{Introduction}

Three keywords in High-Performance Computing (HPC) are Performance, Portability and Productivity, known as the three ``P''. 
Performance is commonly the main focus of Lattice QCD applications, which achieve scalability and high-performance on major computing architectures.
Portability and productivity, instead, are new emerging requirements, usually sacrificed at the cost of performance and fast development, but recently widely chased in response to the new challenges arising in HPC.
In this work, we introduce the Lyncs-API, a new community effort whose goal is to achieve the three ``P'' by developing a Python framework for Lattice QCD calculations.
In particular, we aim to portability and productivity by designing a framework that, respectively, (i) can achieve high-performance on the largest HPC systems and (ii) is flexible, user-friendly, with a clean API and can be easily integrated with other tools as well.

\subsection{Related works in Lattice QCD}

Recently, significant efforts towards portability are being made by developers of various libraries for Lattice QCD. For example, the QUDA library~\cite{Clark:2009wm} is implementing various backends for extending its employment on GPUs made by other manufacturers than NVIDIA, such as AMD or Intel~\cite{quda_talk}. And the Grid library, whose development focused on achieving portability on CPU architectures~\cite{Boyle:2015tjk}, has been also implementing GPU support~\cite{osti_1424990}. 
There are a handful of more examples where the lattice QCD community is either extending or developing new software for adapting to the diversity of computing architectures available nowadays. 
The Lyncs-API is in its way different from these efforts by not implementing directly Lattice QCD kernels but relying on the availability of libraries for providing high-coverage on tools, portability and performance. Indeed, one of its main goals is to bring several widely used libraries for Lattice QCD under a common framework and involve a large community.

On the other hand, productivity is probably the least satisfied feature by Lattice QCD software and it is also the most challenging to achieve. Indeed, most libraries often lack of user-friendliness, flexibility and backwards-compatibility making them difficult to be integrated in applications and also interact between each other. The Lyncs-API, instead, has productivity as its main focus. It is the reason behind the choice of Python as high-level programming language and most of its features are inspired by the impressive achievements of the Python community in this direction. A compatible effort is being made by the Grid Python Toolkit (GPT)~\cite{GPT} which provides a Python interface and Python bindings for Grid. We foresee the usage of GPT in the future for interfacing to the Grid library.

\subsection{Overview of related Python packages}

In recent years, Python has become the most used programming language in data science. This is thanks to its strong community and many successful applications, e.g. in machine learning. The employment of Python in HPC, instead, is  limited but growing and there is a number of tools meant to support it. In the following we will provide an brief overview of key packages used by the Lyncs-API, which facilitate the usage of Python in HPC.

\begin{description}
	\item[\package{Numpy}]\cite{harris2020array} is the most well-known Python package. It offers performant implementations of multi-dimensional array operations and it defines a standard employed by many other packages.
	\item[\package{CuPy}]\cite{Okuta2017CuPyA} is an implementation of Numpy-compatible multi-dimensional array operations on GPUs using CUDA. It supports multiple devices and Just-In-Time (JIT) compilation of kernels. Experimentally, it also supports AMD GPUs using ROCm.
	\item[\package{cppyy}]\cite{lavrijsen2016high} is an automatic, run-time, Python-C++ bindings generator, for calling C/C++ from Python and Python from C++. Libraries are required to be compiled as shared objects and bindings are generated automatically by parsing the header files. The run-time generation enables detailed specialization for higher performance, lazy loading, Python-side cross-inheritance and callbacks for working with C++ frameworks (e.g. template instantiation).
	\item[\package{mpi4py}]\cite{DALCIN20051108} (MPI for Python) provides Python bindings for the Message Passing Interface (MPI) standard with an object oriented interface resembling MPI-2 C++. Since version 3.1.0, it supports GPUDirect RDMA (GDR) communications with GPU-aware MPI implementations.
	\item[\package{h5py}]\cite{collette2013python} implements a Python interface for the HDF5 library and file format. It efficiently supports all features of the HDF5 standard, including parallel IO.
	\item[\package{Dask}]\cite{Dask} provides a Python API for data- and task-parallelism. It supports distributed Numpy arrays via domain decomposition and uses a centralized scheduler for managing multiple workers. Another  focus of the package is remote data representation via \textit{futures}.
	\item[\package{Numba}]\cite{10.1145/2833157.2833162} is an open source JIT compiler that translates a subset of Python and NumPy code into performant machine code. It supports various CPU and GPU architectures and, respectively, \pkg{Numpy} and \pkg{CuPy} arrays. It also supports caching or Ahead-of-Time (AOT) compilation for avoiding repeated overheads during run-time.
\end{description}

\section{Overview of the Lyncs-API}

In the following we provide an overview of the Lyncs-API, focusing on its scope, design and structure. The package as a whole is currently under development and some features described below might be missing or incomplete. Others instead have reached a stable state. In this section we will not discuss implementation status but we will provide a time-agnostic description of the project valid in the future. Implementation details are given in the following section.

\subsection{Statement of purpose}

The Lyncs-API aims to provide a complete framework for performing Lattice QCD calculations using the Python programming language.
The project is community-oriented, open source and welcomes contributions.
Its motivations and purpose are the following:
\begin{itemize}
	\item \textbf{To create a productive, user-friendly and ease-to-use framework for Lattice QCD calculations.} Python is chosen as top-level programming language for implementing the API and it is used for driving and deploying calculations. All computational kernels, instead, are executed by compiled software and libraries developed in C/C++.
	\item \textbf{To achieve performance and portability on HPC computing architectures by interfacing to major libraries for Lattice QCD.} Many libraries are targeted in order to reach high-coverage on tools and performance-portability. Moreover many implementations of the same kernels allow for crosschecks and performance benchmarks. Interfaces to open-source and free-to-use libraries are welcome and the project aims to involve the whole community.
	\item \textbf{To target multitasking, modular computing and resiliency.} The high-level framework is designed to drive calculation on vast partitions of HPC systems beyond scalability of single kernels. This is done via multitasking as a mean to seamlessly increase scalability of applications, to deploy computations simultaneously on partitions with different architectures, i.e. modular computing, and to allow for calculations to proceed in the event of node failure having more nodes at disposal and data duplicates, i.e. resiliency.
	\item \textbf{To follow high-standards in community software development.} We employ version control, unit-testing targeting high-coverage, backwards compatibility, linting, code formatting, issues tracking and updated documentation. We publish on GitHub and use GitHub Actions for Continuous Integration and Deployment (CI/CD). Packages are distributed on the Python Packaging Index (PyPI) and can be installed via \texttt{pip}. Code coverage is tracked in the implementation and new tests are implemented for new functionalities.
\end{itemize}

\subsection{Design and structure}

The Lyncs-API is designed upon a modular structure in order to separate concerns of different modules and maintain their implementation and deployment independent from each other as much as possible. All modules together aim to provide a complete Python framework for Lattice QCD calculations but, on the other hand, they can be used separately within other Python projects allowing for easy code reuse. The modules are developed and deployed independently. We use version control for tracking changes in the software stack and update setup requirements appropriately to ensure that packages point to the correct versions of their dependencies. All packages are collected on the \fhref{https://github.com/Lyncs-API}{GitHub organization}
 with description and documentation in the respective \texttt{README} files and overview on the \fhref{https://lyncs-api.github.io}{online webpage}.

The current structure of packages in the Lyncs-API can be divided in three categories: (i) generic purpose tools, (ii) interfaces to Lattice QCD libraries, and (iii) high-level tools for Lattice QCD calculations. In the first category we find a variety of generic purposes packages that could be also used outside Lattice QCD applications. In the second category we have a number of interfaces to Lattice QCD libraries providing a ``Pythonization'' of their functions/classes resembling as close as possible the library itself. And in the last category we find high-level modules for lattice QCD calculations that provide end-user functionalities based on the low-level packages in the previous categories. More details on the package structure are given in the following sub-sections.

\subsubsection{Generic purpose tools}

Part of the Lyncs-API are a number of packages that provide generic purpose tools whose development has been needed in constructing the API itself. Briefly, we have (i) \lyncs{setuptools}, which provides setup-level functionalities and developer tools including support for CMake; (ii) \lyncs{utils}, which is a collection of small and independent functionalities that recur widely in the implementation of all packages; (iii) \lyncs{cppyy}, which provides a wrapping layer on top of \pkg{cppyy} for simplifying its usage when interfacing to libraries (more details in Section~\ref{sec:cppyy}); (iv) \lyncs{mpi} with tools for MPI using \pkg{mpi4py} and in combination with \pkg{Dask}; and lastly we have two major packages, \texttt{lyncs\_io} and \texttt{lyncs\_field}, described in the following.

\lyncs{io} aims to provide an all-inclusive framework for IO in Python. It responds to the necessity of missing tools for parallel IO and user-friendliness in various file formats. Therefore it implements a very simple interface with two main functionalities, \texttt{load} and \texttt{save}; it supports various file formats and, where possible, parallel IO for domain decomposed arrays. Additionally, it also implements file formats that are common in Lattice QCD as, for instance, the Lattice QCD Interchange Message Encapsulation (LIME). More details are given in Section~\ref{sec:IO}.

\lyncs{field}\footnote{currently under development} provides a set of meta-classes for abstraction of properties of multi-dimensional arrays and tensor fields. Indeed, an issue when interfacing to various libraries is the variety of formats and conventions used for storing the data. Especially in Lattice QCD, many choices can be made in ordering the axis, whether using a checkerboard splitting, how to store the halos, etc. With \texttt{lyncs\_field} one can easily track information on the properties of arrays and seamlessly convert from a standard to another. \pkg{Numpy} and \pkg{CuPy} are the default backends for array's operations on CPU and GPU architectures, respectively, but also any Numpy-compatible backend is supported.

\subsubsection{Interfaces to Lattice QCD libraries}

A major task of the Lyncs-API is to provide Python interfaces to a wide range of libraries. Within the scope of the Lyncs-API is to target both, cutting-edge software packages as well as legacy software, in order to achieve on one side high-coverage on tools and performance-portability as well on the other to perform crosschecks and benchmarks ensuring continuity in the developments. Currently the number of interfaces is limited to \lyncs{DDalphaAMG}, \lyncs{quda}, \lyncs{tmLQCD} and \lyncs{clime} that interface respectively to \href{https://github.com/sbacchio/DDalphaAMG}{DDalphaAMG}~\cite{Bacchio:2016bwn}, \href{https://github.com/lattice/quda}{QUDA}~\cite{maddyscientist_2021_5610079}, \href{https://github.com/etmc/tmLQCD}{tmLQCD}~\cite{Jansen:2009xp} and \fhref{https://usqcd-software.github.io/c-lime/}{c-lime}.  The interfaces follow as closely as possible the structure of the libraries in order to simplify their implementation and  to remain familiar to users of the library itself. Major differences may appear when libraries are not object-oriented. We expect in the future to increase significantly the list of targeted libraries while enlarging the community. More details on the interfaces are given in Section~\ref{sec:interface} with examples in Section~\ref{sec:examples}.

\subsubsection{High-level tools for Lattice QCD}

Apart from interfaces to libraries, all other tools for Lattice QCD are expected to be on the top level only. Here we find generic APIs, abstraction of common workflows and applications. For instance, we have implemented tools for Hybrid Monte-Carlo (HMC) in \lyncs{hmc} and we aim to provide a complete end-user API in \Lyncs. Unfortunately, this is currently in an early development stage and we do not have a complete list of packages yet.

\section{Implementation details}

As typical when developing high-level software, the design of the Lyncs-API proceeded top-down while its implementation is proceeding bottom-up. For this reason, not all features described in the previous section are available, but, on the other hand, the modular structure of the Lyncs-API allows for most of the developed components to be used as stand-alone packages. Hereafter we provide implementation details on key components with examples.

\subsection{Interfaces to libraries}\label{sec:interface}

All packages that interface to libraries implement the following procedure:
\begin{enumerate}
	\item Library's source-code is cloned during installation from its online repository targeting a chosen release (or \textit{commit}). We fix the latter, because most libraries do not ensure backward compatibility and/or stability. Update of the library's release is done on a regular basis after respectively updating the interface and testing all functionalities.
	\item Patches to the source-code are applied where needed and they are distributed within the interface repository. Patches either fix installation or execution issues, e.g. enabling compilation of the library as shared-object or exposing more features in the interface.
	\item The above items, compilation and installation of the library are managed by CMake. Indeed, libraries are not compiled AOT and are not distributed as Python \textit{wheels} for optimization purposes. The installation of the package, including the execution of CMake, follows the default Python procedure with setup script in \texttt{setup.py} that can be installed via \texttt{pip}.
	\item Python bindings are automatically generated via \pkg{cppyy} using our wrapper \lyncs{cppyy}, which simplifies its usage with some additional features (see \ref{sec:cppyy}).
	\item A uniform and well documented Python interface is created for key features of the library resembling its structure. Data conversion and input validation are also part of this step, using typical Python features in data management and error handling.
	\item Unit-testing and high-coverage of the interfaced functionalities is reached. Functions are tested for many lattice sizes, parallel partitioning, data formats, floating precision, etc. sometimes finding issues and providing feedbacks for the developers of the library itself.
\end{enumerate}

\subsubsection{Automatic bindings with cppyy}\label{sec:cppyy}
Giving more details on the fourth step, \pkg{cppyy} automatically generates Python bindings for C/C++ libraries by parsing header files and loading the compiled library at run-time as a shared object. By collecting information on the signature of functions and classes in the headers, it can then allocate appropriately memory, manage data, call functions, convert Python objects to inputs and vice versa for outputs. This process is based on Cling~\cite{Vasilev_2012}, an interactive C++ interpreter, built on the top of LLVM and Clang libraries with specific Python functionalities provided by cppyy. The package also supports JIT compilation of C++ templates and headers-only implementations. The standard API of \pkg{cppyy} provides two functions for this purpose \texttt{include} and \texttt{load\_library}. In \lyncs{cppyy} we use an object-oriented approach representing the library as a class. For example,
\begin{lstlisting}
  from lyncs_cppyy import Lib
  lib = Lib(header="quda.h", library="quda", namespace="quda")
\end{lstlisting}
the above can be used to load e.g.~\texttt{quda} and access its functionalities as attributes/methods of the \texttt{lib} object. Additional features compared to \texttt{cppyy} is the usage of a default \textit{namespace}, lazy instantiation (loading the library at its first use), definition and access to macros.
Additional remarks on using \pkg{cppyy} compared other standard approaches are
\begin{itemize}
	\item \textbf{Resiliency to changes:} since interfaces are not compiled AOT, all functionalities not affected by changes are operative and usable. Any issue will appear at run-time. This is on one side an advantage when transiting from a version of a library to another, but on the other interfaces require full coverage for catching any broken features.
	\item \textbf{Error tolerance:} in connection to the above, \pkg{cppyy} is error tolerant and allows for the usage of ``try/except'' when calling functions or instantiating classes. This is useful for supporting multiple versions of libraries and easily addressing changes. Moreover, cppyy can also convert abortive signals, e.g. segmentation faults, into Python exceptions avoiding to interrupt execution and easily debug software.
	\item \textbf{Class cross-inheritance:} another very useful feature is cross-inheritance of C++ classes in Python allowing the latter to be called by C++. Indeed, method overloading and virtual method overriding are fully supported.
\end{itemize}

\subsubsection{Examples}\label{sec:examples}

Hereafter we provide two examples for computing the inverse of the Dirac operator using Lyncs' interfaces.

\noindent
\begin{minipage}{0.5\linewidth}
\begin{lstlisting}
 import lyncs_quda as quda
 from mpi4py import MPI

 # Creating a Cart. comm via MPI
 lattice = [4, 4, 4, 4]
 procs = [2, 2, 1, 1]
 comm = MPI.COMM_WORLD
 comm = comm.Create_cart(procs)

 # Creating a random gauge field
 gauge = quda.gauge(lattice=lattice,
            comm=comm, device="GPU")
 gauge.random()
 plaq = gauge.plaquette
 print("Plaquette:", plaq)
 
 # Creating a random spinor
 rhs = quda.spinor_like(gauge)
 rhs.random()
 
 # Inverting using CG
 dirac = gauge.Dirac(kappa=0.125)
 mat = dirac.MMdag
 sol = mat.solve(rhs, inv_type="CG")
 res = mat(sol) - rhs
\end{lstlisting}
\end{minipage}
\begin{minipage}{0.5\linewidth}
\begin{lstlisting}
 from lyncs_DDalphaAMG import Solver
 from lyncs_mpi import Client

 # Creating a Cart. comm via Dask
 lattice = [4, 4, 4, 4]
 procs = [2, 2, 1, 1]
 client = Client()
 comm = client.create_cart(procs)

 # Instantiating DDalphaAMG
 solver = Solver(lattice=lattice,
         comm=comm, kappa=0.125)

 # Reading a random configurations
 conf = solver.read_configuration(
         "test/conf.random")
 plaq = solver.set_configuration(
         conf)
 print("Plaquette:", plaq)

 # Inverting using multigrid
 rhs = solver.random()
 sol = solver.solve(rhs)
 res = solver.D(sol) - rhs
 res = res.compute()
\end{lstlisting}
\end{minipage}

\noindent
The first example (left panel) uses QUDA via \lyncs{quda} and the second uses DDalphaAMG via \lyncs{DDalphaAMG}. Notation in the two implementation differ slightly since the interfaces follow closesly the structure of the libraries. The interface to QUDA is based on its C++ implementation (not the C interface) and the interface to DDalphaAMG is based on the header \fhref{https://github.com/sbacchio/DDalphaAMG/blob/master/src/DDalphaAMG.h}{\texttt{DDalphaAMG.h}}. Additionally the first example parallelizes directly via MPI while the second uses Dask at the higher level which then uses MPI. The two approaches are interchangeable and they have a very similar initialization (i.e. need to exchange how the communicator is created in the two examples). On the other hand significant differences in the execution apply as outlined in the following section.

\subsection{Parallel computing with Dask and MPI}

Beside standard data-parallelism via domain decomposition using MPI, the Lyncs-API also employs \pkg{Dask} for more advanced features, e.g. task parallelism. In Lyncs the usage of Dask is optional, made compatible with MPI and ease to use (see above example). The core of Dask is a centralized scheduler that receives instructions from a client and assigns tasks to workers. Within the duties of the scheduler are knowledge on the data location, properties and status of workers, management and transfer of data, submission of tasks and dependencies resolution. On top of that, Dask provides high-level interfaces like distributed Numpy arrays via domain decomposition which is of interest for Lattice QCD and it supports all features needed including \textit{halos}. In \lyncs{mpi} we have developed tools for bridging the gap between Dask and MPI, e.g. grouping workers within a communicator and easy integration in classes.

Going into details of the above examples, while in the left panel the computation proceeds as expected in a MPI-distributed application, in the right panel most of the calculations are delayed. Starting from the creation of the communicator, nothing is actually computed until ``\texttt{.compute()}'' is called. This means that all steps are collected in a computational graph, which is then passed to the scheduler and computed by the first group of ``workers'' available.

\subsection{Simplified parallel IO in Python}\label{sec:IO}

Another outcome of the Lyncs-API is a high-level tool for parallel I/O in Python within \lyncs{io}. The package has a general purpose and implements a simple API for loading and saving distributed data using domain decomposition via either Dask or an MPI Cartesian communicator. The objectives of the package are

\begin{itemize}
	\item \textbf{Seamlessly IO:} we have tried to simplify the IO as much as possible, without having to open files, directly specify the format or using more than one line of code. Three main functions are provided \texttt{save(obj, filename)} for saving,
	\texttt{obj=load(filename)} for loading and \texttt{obj=head(filename)} for retrieving metadata without loading data.
	More arguments are available as outlined in the following.
	\item \textbf{Many formats supported}. The file format can be either deduced
	from the \texttt{filename}'s extension or with the option \texttt{format} passed as an argument to the above functions.
	 The structure of the package is flexible enough to
	easily accommodate new/customized file formats as these arise.
	For instance we support file formats typical in lattice QCD like LIME.
	\item \textbf{Omission of extension} When saving, if the extension is omitted,
	the optimal file format is deduced from the data type and the extension
	is added to the filename. When loading, any extension is considered,
	i.e. \texttt{filename.*}, and if only one match is available, the file is loaded.
	\item \textbf{Support for archives}. In case of archives, e.g. HDF5, zip etc.,
	the content can be accessed directly by specifying it in the path.
	For instance using \texttt{directory/file.h5/content}, \texttt{directory/file.h5}
	is the file path, and the remaining is content to be accessed that
	will be searched inside the file (this is inspired by \pkg{h5py}).
	\item \textbf{Support for Parallel IO}. Where possible and implemented,
	parallel IO is supported. This is enabled either via MPI providing
	a Cartesian communicator with the option \texttt{comm}, or via \pkg{Dask}
	with the option \texttt{chunks}.
\end{itemize}

\section{Conclusions and future plans}

In this contribution we have presented the pillars of the Lyncs-API, a new software effort for Lattice QCD applications using the Python programming language. We have described its design, purpose and long term vision. We have provided details on the implementation of its low-level building blocks, such as interfaces to libraries, data- and task-parallelism and I/O. We have motivated the choice of Python as a tool for easily addressing new computational challenges in HPC and Lattice QCD, such as the integration of machine learning techniques, usage of task parallelism and achieving the three-Ps towards Exascale supercomputing. The development of the Lyncs-API is strongly motivated by these and wants to be an alternative to more traditional/low-level approaches.

In the short term, our plan is to complete the interface to QUDA, \lyncs{quda}, closely following the developments of the library towards 2.0, and to implement \lyncs{field} which will be the connecting layer between all libraries allowing for easy exchange and reordering of data. This will be followed by the implementation of more high-level tools and first large-scale applications. A major target of our development is to achieve scalability in HMC simulations using modular computing and multi-tasking.
Concluding, with this presentation we wanted to raise awareness on the project and welcome contributions and/or feedback.

\section*{Acknowledgments}

This project has received funding under PRACE-6IP, Grant agreement ID: 823767, Project name: LyNcs. LyNcs is one of 10 applications supported by PRACE-6IP, WP8 ``Forward Looking Software Solutions''. S.B., J.F. and C.S. have received funding under this project. The authors would also like to thank the other members of the LyNcs project for the stimulating collaboration.

\bibliographystyle{JHEP}
\bibliography{refs}

\end{document}